\documentclass{aastex}

\slugcomment{Submitted to ApJL}
\shorttitle{The {\it Forbidden} Abundance of Oxygen}
\shortauthors{Allende Prieto, Lambert \& Asplund}

\begin{document}

\title{The {\it Forbidden} Abundance of Oxygen in the Sun}

\author{Carlos Allende Prieto and David L. Lambert}
\affil{McDonald Observatory and Department of Astronomy, The University 
of Texas, Austin, TX 78712-1083, USA}

\and

\author{Martin Asplund}
\affil{Uppsala Astronomiska Observatorium, Box 515, Uppsala 75120, SWEDEN}

\begin{abstract}

We reexamine closely the solar
photospheric line at 6300 \AA, which is attributed to  a
 forbidden
line of neutral oxygen, and is widely used in analyses of other late-type stars.
 We use a three-dimensional
time-dependent hydrodynamical model solar atmosphere which
has been tested successfully against observed granulation patterns
and an array of absorption lines. We show that the solar line is a blend with a
Ni\,{\sc i} line, as previously suggested but oftentimes neglected.
Thanks to accurate atomic data on the [O\,{\sc i}] and Ni\,{\sc i} lines
we are able to derive an accurate oxygen abundance for the Sun:
 $\log \epsilon ({\rm O}) = 8.69 \pm 0.05$ dex, a value  at the lower
end of the distribution of previously published abundances, but in 
good agreement
with    estimates  for the local interstellar medium and hot
stars in the solar neighborhood. We conclude by discussing the
 implication of the Ni\,{\sc i} blend on oxygen abundances derived from the
[O\,{\sc i}] 6300 \AA\ line in disk and halo stars.

\end{abstract}



\section{Introduction}

	The cosmic abundance of oxygen is an important number in a wide
variety of scenarios.  Several debates have over the years
focussed on disagreements about the oxygen abundance of stars, and
disparities between estimated abundances for different kinds of
astronomical objects.
For example, there is a long history of using
 hot stars as tracers of the galactic radial variation of
the oxygen  abundance, but only very recently  have the stellar abundances
seemed to agree with the results derived from emission lines in H\,{\sc ii}
 regions  (Rolleston et al. 2000). Currently, the study of metal-poor stars
is particularly polemical. 
 A group of astronomers  defends a monotonic
increase of the oxygen-to-iron ratios (i.e., [O/Fe])
 with lower iron abundances
(see, e.g., Israelian et al. 2001; Boesgaard et al. 1999), 
whereas a second group
tends to prefer an essentially constant [O/Fe] for stars with [Fe/H] $ \lesssim -1$
(see, e.g., Carretta, Gratton \& Sneden 2000;  
Nissen, Primas \& Asplund 2000). 
The discrepancy for metal deficient stars may be connected to the
use of different indicators of the oxygen abundances and, ultimately,
caused by deficiencies in the modeling of the spectra. 
 
The solar oxygen abundance is not free from debate. In fact, the
 solar controversy and that in 
metal-poor stars could be related. 
The most popular indices of the oxygen abundance in cool stars
 are i) the  [O\,{\sc i}] forbidden lines at 
6300 and 6363 \AA; ii) the OH lines in the UV and IR; and 
iii) the O\,{\sc i} triplet at about 7773 \AA\footnote{Unfortunately,  
 weaker (forbidden and allowed) O\,{\sc i} lines in 
the  optical and near-IR solar spectrum appear to be either heavily blended 
with several (some times unknown) other features (e.g. 5577.3, 6156.8, 
6158.2, 9741.5, 11302.2 \AA), lack atomic data (e.g. 9760.7 \AA),or  are 
affected by  a  
large uncertainty in the continuum location (e.g. 8446.3, 9265.99 \AA), 
or suffer some of these problems 
simultaneously, occasionally in combination with serious 
departures from LTE (e.g. 8446.8 \AA).}. In the solar case, the forbidden 
 lines and the OH lines tend to provide higher abundances, by about $0.1-0.2$ 
 dex, than the O\,{\sc i} 7773\AA\  triplet,  
 when allowance for departures from Local Thermodynamic Equilibrium (LTE) 
 in the formation of the triplet is made. 

  None of the indicators of the solar oxygen abundance  is free from suspicion.
 The  popular
[O\,{\sc i}] lines are threatened by 
the presence of blends (Lambert 1978). The O\,{\sc i} 7773 \AA\ triplet 
likely suffers important departures 
from  LTE (see e.g. Kiselman 1993; 
Takeda 1994; Reetz 1999a,b). The OH lines are 
particularly sensitive to temperature, and 
therefore compromised by deficiencies in the model atmospheres (e.g., 
Sauval et al. 1984; Asplund \& Garc\'{\i}a P\'erez 2001).  In addition,  
 all lines are affected  by surface inhomogeneities (granulation) 
  that are ignored by classical model
atmospheres (Kiselman \& Nordlund 1995).  Some lines are likely 
afflicted by several errors acting in concert, for example,  the OH UV lines 
may be affected not only by model atmosphere uncertainties but also
 by significant  
departures from LTE, and the uncertainties 
in the opacities in that spectral region (Hinkle \& Lambert 1975; Bell, 
Balachandran \& Bautista 2001; Asplund \& Garc\'{\i}a  P\'erez 2001). 

We reexamine the solar [O\,{\sc i}] 6300 \AA\ line
using synthetic spectra based on a three-dimensional hydrodynamical
model atmosphere, and allowing for  
the Ni\,{\sc i} blend. A principal advantage of the [O\,{\sc i}] line
is that departures from LTE should be very small. A disadvantage is that
the line is weak and, therefore, susceptible to blends.
As discussed by Lambert (1978),  the line at 6300  \AA\ is blended
with a Ni\,{\sc i} feature, while that at 6363 \AA\ is affected by a CN line. 
We consider only the 6300 \AA\ line because the blending Ni\,{\sc i}
line makes a smaller contribution to the solar line than does the
CN line to the 6363 \AA\ line (Reetz 1999a), the 6363 line is located in 
the middle of a broad Ca\,{\sc i} auto-ionization line and, as discussed for OH,  
molecular features are particularly sensitive to the details of the 
temperature structure. The next section describes the basic ingredients 
involved 
in the  analysis, \S 3 is devoted to the comparison with the solar flux 
spectrum  
to derive the photospheric oxygen abundance, and \S 4 is a brief
summary of the results.

\section{Model atmospheres and line synthesis}

We  made use of a 3D time-dependent hydrodynamical simulation of the 
solar surface calculated  by Asplund et al. (2000), and based on the 
compressible radiative hydrodynamical code described by 
Stein \& Nordlund (1998). The equations of mass, momentum, and energy 
conservation
together with the simultaneous treatment of the 3D radiative transfer 
equation were solved on a Eulerian grid with $200\times200\times82$ points 
to represent $(6\times6\times3.8) \times 10^3$ km in the solar surface, with
about $10^3$ km above continuum optical depth unity. For more details
we refer the reader to Asplund et al. (2000) and references therein.

The calculated  flux profiles for the [O\,{\sc i}] and Ni\,{\sc i}
 lines at 6300 \AA\ from
the 3D simulation are based on a sequence of 50 minutes, taking a snapshot 
every 30 seconds. The integration over the disk used $4\times4$ angles, 
and took into account the solar surface rotation as a solid body 
($v_{\rm rot} \sin i \simeq 1.9$ km s$^{-1}$). The Uppsala synthesis package
(Gustafsson et al. 1975 with subsequent updates) was the source of the continuum opacities, 
partition functions, ionization potential and other basic data for the line
synthesis as well as for the simulation. Collisional broadening by hydrogen 
atoms was considered following the van der Waals formula, but is essentially
irrelevant
for the extremely weak lines considered (equivalent widths
$\lesssim$ 5 m\AA). Natural broadening, also irrelevant,
 was included, using data from
the Vienna Atomic Line Database (VALD; Kupka et al. 1999). 
Reduction of the partial pressure of oxygen through CO formation is
included with the carbon abundance put at $\log \epsilon({\rm C})$ = 8.52;
though the effect on the number density of O\,{\sc i} is $\le 5$\%. 
Our assumption of LTE for the [O\,{\sc i}] line formation is supported 
by our own NLTE calculation for various 1D model solar atmospheres.
It must be stressed that the line profiles are predicted {\it ab initio},
i.e., microturbulence and macroturbulence are not invoked and adjusted
to fit these or other lines. Convective velocity fields 
and thermal broadening are provided by the model.

Our goal was to determine the best fit to the solar 6300 \AA\ feature using
the known atomic parameters for the [O\,{\sc i}] and Ni\,{\sc i}
lines. The laboratory wavelength of the [O\,{\sc i}] line was measured by
Eriksson (1965) as $6300.304 \pm 0.002$ \AA. 
The transition probability for the [O\,{\sc i}] line is dominated
by the magnetic dipole contribution  with a minor (0.3\%) 
electric quadrupole contribution.  We adopt 
Storey \&  Zeippen's  (2000) relativistic  calculation of 
the magnetic dipole contribution, and following them include the
electric quadrupole contribution from Galav\'{\i}s, Mendoza, \&
Zeippen (1997) to obtain $\log gf = -9.717$ with an accuracy of
a few per cent.  Inclusion of relativistic corrections reduced the value
by less than 2 \%. The  $\log gf$  has changed little over the years; 
Lambert (1978)  suggested  $-9.75$, after comparison of laboratory
and older theoretical determinations.

The blending Ni\,{\sc i} line's wavelength was measured as 6300.339 \AA\ by
Litz\'{e}n, Brault, \& Thorne 
 (1993) with an estimated uncertainty of 2 to 3 m\AA. Litz\'{e}n et al.'s
energy levels derived from their extensive reinvestigation of the Ni\,{\sc i}
spectrum predict 6300.342 \AA\ for the line. 
The NIST web database gives 6300.343 \AA, as 
calculated from the energy levels of Sugar \& Corliss (1985).
Litz\'{e}n et al. did not report measurable isotopic splitting and, therefore,
we neglect isotopic (and hyperfine) splitting. 
We have adopted 
6300.339 \AA\ as
the rest wavelength of Ni\,{\sc i} line. 
The $\log gf$ for the Ni I line is uncertain. There are seemingly no
laboratory measurements for this line. The energy levels clearly show that
LS-coupling is an inappropriate assumption from which to derive the
solar $gf$-value using Ni\,{\sc i} lines of the same and similar
multiplets. Indeed, Moore's (1959)  classification gives the transition
as y\,$^3$D$^\circ_1$ - e\,$^3$P$_0$ but severe configuration interaction
renders these misleading labels. Litz\'{e}n et al. show that the
dominant (90\%) contribution to the upper term is 
not from the $^3$P term but from a $^1$S term, and while the leading
 contribution to the lower term is from the $^3$D term there is
an appreciable contribution (34\%) from other terms. 
These contributions differ for different levels of the lower and upper
terms.
We treat the product $gf \epsilon$({\rm Ni}) as a free parameter.

\section{Comparison with the observed spectrum}

We have confronted our calculated profiles with the  FTS solar flux 
spectrum published by Kurucz et al. (1984). The resolving power is about
half a million, and the signal-to-noise ratio amounts to several thousand, 
making these data adequate for studying the [O\,{\sc i}] feature at 6300 \AA. Allende
Prieto \& Garc\'{\i}a L\'opez (1998) checked that the wavelength scale of the
atlas is very accurate. It has been corrected for the Earth-Sun velocity,
 but not
for the gravitational redshift, which amounts to 633 m s$^{-1}$ for the
photospheric spectrum intercepted at Earth.

Two strong lines (Si\,{\sc i} 6299.6 \AA\ and Fe\,{\sc i}
6301.5 \AA) depress the continuum in the immediate  vicinity
of the [O\,{\sc i}] - Ni\,{\sc i} blend,
 but have a negligible effect on the shape of the
observed blend. Other weaker lines might be present closer to the blend,
as reflected by the  slightly irregular shape of the spectrum, 
discouraging the  individual modeling of all lines, but rather
suggesting a zeroth-order correction to the continuum level, which
becomes one of the fitting parameters.  As the 3D models predict
the convective line asymmetries and shifts for weak neutral lines to well
within 0.1 km s$^{-1}$ ($\approx 0.002$ \AA;  Asplund et al. 2000), 
the  shifts - absolute and differential - of the oxygen and nickel lines 
from the  laboratory wavelengths cannot
be adjusted. Therefore, the only  parameters we can tune to match the
predicted profile to the observed profile are the continuum
level, the oxygen abundance, and the  product $gf 
\epsilon$({\rm Ni}).  The optimization of the fit was accomplished using
the slow-but-robust amoeba algorithm based on the  Nelder-Mead simplex
method (Nelder \& Mead 1965; Press et al. 1986).

\begin{figure}
\epsscale{0.7}
\plotone{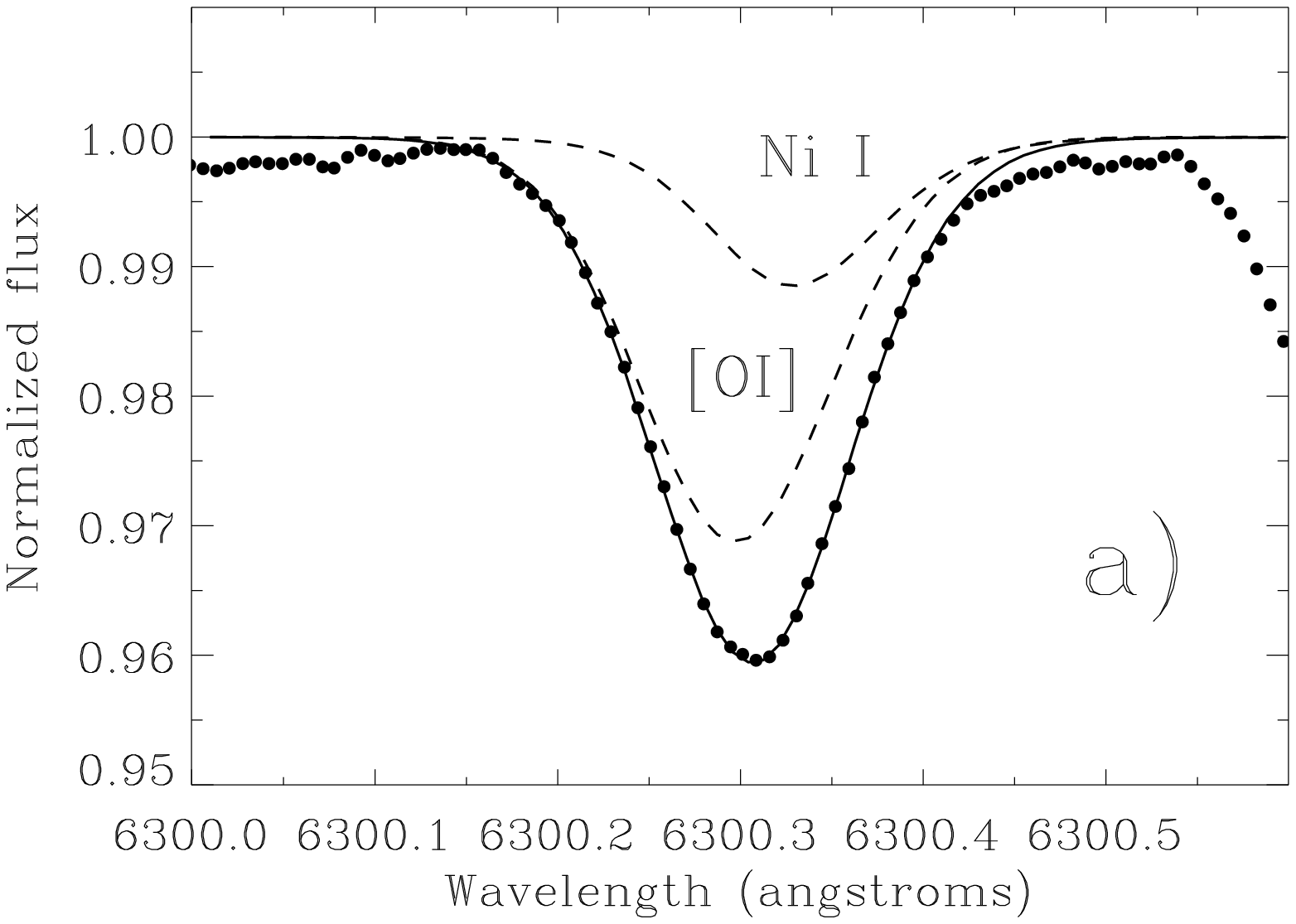}
\plotone{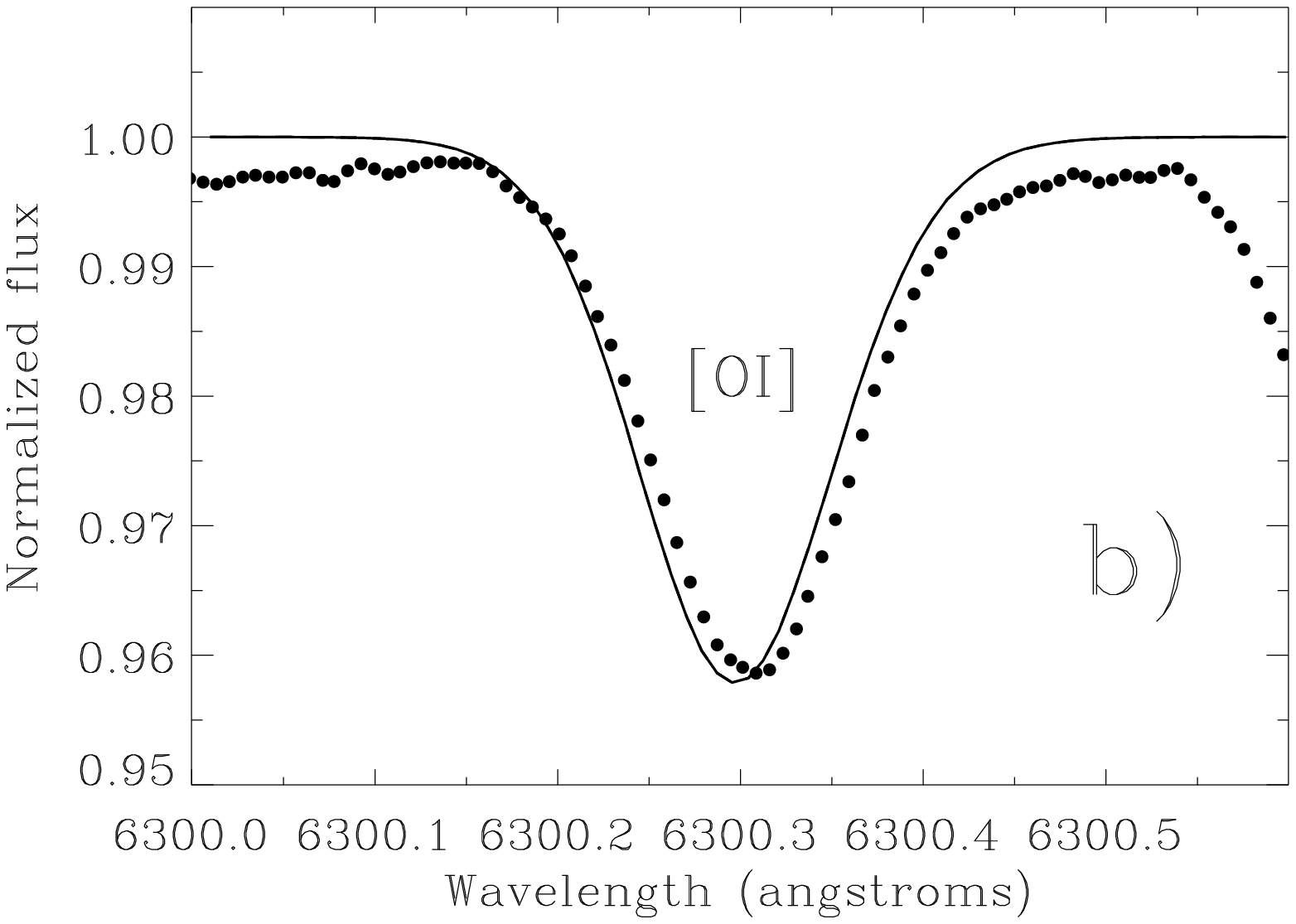}
\figcaption{a) Comparison between the observed (filled circles) 
and synthetic (solid line)
profiles after the $\chi^2$ minimization. The individual calculations of 
the oxygen and nickel lines 
are also shown with dashed lines.  b) Best fit assuming  
the observed feature is entirely produced
by the  oxygen forbidden line.  
\label{3d}} 
\end{figure}

Figure \ref{3d}a shows the agreement between the observed profile 
(filled circles) and that calculated with the 3D  
model atmosphere (solid line). The oxygen and nickel lines are represented 
with dashed lines.  Adopting
a signal-to-noise-ratio (S/N) of 2100, estimated from the region between
6303.10 and 6303.25 \AA, we choose to minimize the reduced $\chi^2$ in the
region between  6300.15--6300.40 \AA\ to avoid apparent blends outside 
the selected region, which  
leads to a multiplicative correction to the  local continuum level 
in the solar atlas of $C=1.0083$. 
Alternative reasonable choices of $C$ 
do not change the retrieved oxygen
abundance by more than 0.02 dex (see below). 
A change as small as 0.005
\AA\ in any direction in the wavelengths of the oxygen or nickel 
lines would degrade  the final $\chi^2$ of an optimal fit. 

 The reduced
$\chi^2$ (34 frequencies and 3 degrees of freedom) for the flux between
6300.15 and  6300.40 \AA\ is 0.30, and the probability for this  
to happen by chance is $P = 5 \times 10^{-5}$. 
We stress that we have corrected  for
 the Sun-Earth gravitational redshift (633 m
s$^{-1}$), but no arbitrary shifts have been allowed for the oxygen,
nor for the nickel lines, nor for both as a pair. No macro-turbulence
or micro-turbulence enter the fit. The effective temperature or,
equivalently, the entropy at the lower boundary of the simulation box,
was carefully adjusted prior to starting the simulation. The temporal 
average of $T_{\rm eff}$ from the simulation is $5767 \pm 21$ K, where the 
quoted error is the standard deviation for the 100 snapshots. The 
adopted surface gravity is $\log g = 4.437$ (cgs). The uncertainties
  in the solar gravity, effective temperature, and chemical composition  have a
negligible impact on the fit.

External uncertainties inherent in the adopted
set of continuum opacities, partition functions, equation of
state, etc. are considered as part of our systematic errors. From tests with 
different synthesis codes and previous experience,
we expect these not to exceed 0.02 dex. The estimated 
uncertainty in the theoretical value adopted for the transition 
probability of the forbidden line adds a contribution of similar size. 
An uncertainty in the wavelength of the Ni\,{\sc i} of $\pm 0.003$ \AA\ 
translates to an error in the oxygen abundance of $^{+0.02}_{-0.02}$ dex 
and $^{-0.03}_{+0.04}$ in 
$\log [gf \epsilon({\rm Ni})]$. The same shifts in the wavelength of the 
[OI] line would produce a change of $^{+0.04}_{-0.03}$ dex 
in the oxygen abundance  and $^{-0.15}_{+0.10}$ dex in 
$\log [gf \epsilon({\rm Ni})]$. 
Simultaneous shift of the rest central wavelengths of both lines by 
$\pm 0.003$ \AA\ would induce  
changes to the determined abundance of oxygen by  $^{+0.04}_{-0.05}$ dex, and 
$^{-0.18}_{+0.12}$ dex in $\log [gf \epsilon({\rm Ni})]$.

\begin{figure}
\epsscale{0.8}
\plotone{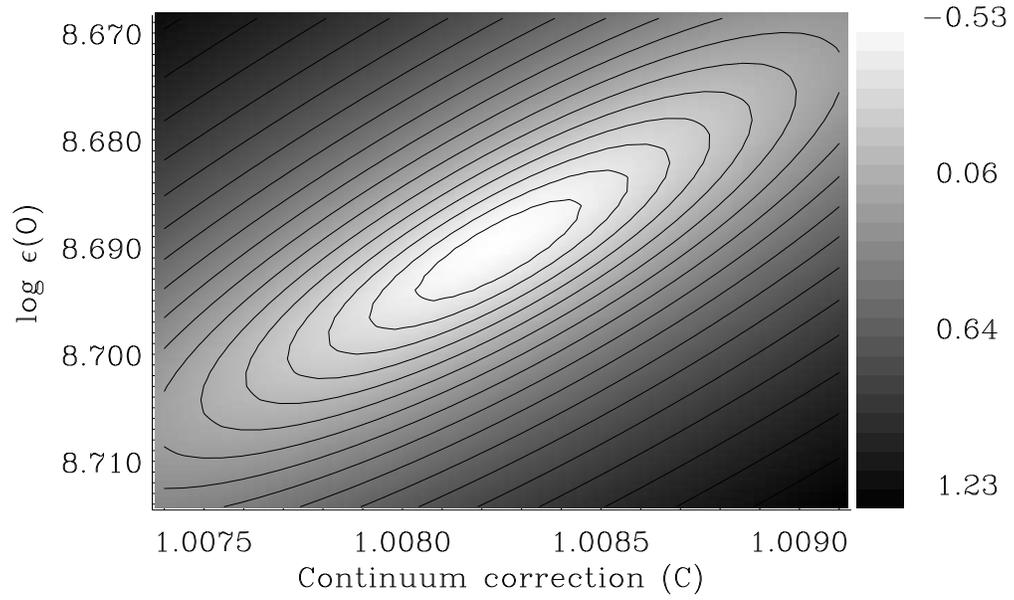} 
\figcaption{Contour and gray-scale plot showing the variation of the reduced
$\chi^2$ as a function of the continuum level and the oxygen abundance, while keeping
the nickel abundance constant. The plot uses logarithmic units, and therefore
the units in the code bar of the  gray-scale are dex.
\label{chi}} 
\end{figure}

 We find $\log \epsilon ({\rm O}) = 8.69 \pm 0.05$ dex.   
 Fig. \ref{chi} shows the projection onto the
$C-\log \epsilon({\rm O})$ plane of the reduced $\chi^2$. The scale for both
the gray-scale and the contours is logarithmic and, therefore, the
units in the gray-scale code bar are dex. 

 Our evaluation of the Ni\,{\sc i} line 
is equivalent to  $\log [gf
\epsilon({\rm Ni})]=3.94$. Adopting the solar Ni abundance from Grevesse \&
Sauval (1998), we obtain $\log gf$ = -2.31 for the Ni\,{\sc i} line, which
compares with $\log gf$ = -1.74 computed by Kurucz \& Bell (1995).
 Figure \ref{3d}b shows the
best fit obtained assuming the solar feature at 6300 \AA\ is entirely
produced by the oxygen forbidden line: the abundance derived would be
$\log \epsilon ({\rm O}) = 8.82 $ dex, which allowing for the differences 
in the adopted $\log gf$s and the correction due to consideration of 3D
($\simeq 0.08$ dex),  is in perfect agreement with previous
results  based on 1D model atmospheres and neglect of the Ni\,{\sc i}
blending line. If a redshift of about 10 m\AA\ (476 m s$^{-1}$) is applied to the
predicted profile, the fit to the observations is obviously improved, but
we argue that such a shift far exceeds known sources of wavelength-related
errors. 

Unpublished measurements of the  Ni\,{\sc i} line 
report the detection of five isotopic components (referee's private 
communication). Adopting those isotopic shifts and solar system abundance 
ratios (Anders \& Grevesse 1989), we find a slightly worse fit to the 
observed feature (reduced $\chi^2 = 0.45$),  and our estimate of the solar 
oxygen abundance 
increases by 0.01 dex. Kurucz lists 70  weak CN lines between 6300.0 
and 6300.6 \AA\footnote{http://kurucz.harvard.edu}, which should be responsible 
for at least part of the continuum depression that we find. A significant 
unrecognized blend or blends in the observed feature would  further 
lower the oxygen abundance, although the extremely good fit that we 
find  constrains tightly such possibility.

Reetz (1999a,) analyzed the 6300 \AA\ line using empirical
1D model atmospheres and high S/N spectra of the
solar disk-center and limb. The observed profile was fitted not only
by adjusting the contributions of the [O\,{\sc i}] and Ni\,{\sc i} lines
but also by including microturbulence and macroturbulence. 
With the Holweger-M\"{u}ller (1974) atmosphere, the oxygen abundance including
a Ni\,{\sc i} contribution, Reetz obtained $\log \epsilon ({\rm O}) = 8.81$
from fits to the center and the limb spectra. His value
on correction to our adopted $gf$-value for the [O\,{\sc i}] line
corresponds to $\log \epsilon ({\rm O})$ = 8.75, a difference of 0.06 dex from
our result. This difference could be bridged by adopting another
choice for the empirical model -- see Reetz (1999a, Table 3.3).
The Ni\,{\sc i} line's $gf$-value was set by considering other but similar
lines and the adjustments to the Kurucz $gf$-values needed to fit the
solar lines. 
Reetz apparently fitted the 6300 \AA\ line profile equally well with and
without the Ni\,{\sc i} line and, therefore, his preferred lower oxygen
abundance is based on the argument used to set the Ni $gf$-value 
($\log gf = -1.95$). In our
case, the line profile is used to set the Ni\,{\sc i} line's
contribution.

\section{Summary and conclusions}

The photospheric 
oxygen abundance we determine from the 6300
\AA\ feature, $\log \epsilon ({\rm O}) = 8.69 \pm 0.05$ dex, is at the
lower end of the distribution of oxygen abundances previously published.
 This lower 
value results  partly from adoption of the recent relativistic calculation
of the transition probability, but is primarily 
 the result of two factors: (i) the use of the 3D hydrodynamical model
atmosphere instead of the Holweger-M\"{u}ller model ($\simeq -0.08$ dex), 
and (ii) an accounting for the contribution of the Ni\,{\sc i} line 
($\simeq -0.13$ dex). The lower oxygen abundance is
consistent with the oxygen abundances found in the local 
interstellar medium when adopting an oxygen gas-to-dust ratio of $\sim 2$
(Meyer, Jura \& Cardelli 1998),  and in the photospheres of hot stars 
in the solar neighborhood (e.g. Cunha \& Lambert 1994; Kilian, 
Montenbruck, \& Nissen 1994), both of 
which typically fall in the range $8.65-8.70$ dex.

As [Ni/Fe] $\simeq 0$, and oxygen is known to be relatively more abundant at lower metallicities,
the importance of the nickel line in the 6300 \AA\ blend will decrease in
metal-poor stars, becoming negligible below [Fe/H] $\simeq -1$. However, 
it is important to note that the common practice of using a scale 
relative to the solar abundances would affect the shape of any possible
 trend of [O/Fe] in the range $-1 \le $ [Fe/H] $ \le 0$. In 
 particular, the nearly constant [O/Fe]  in stars with 
 $-0.3 \leq $ [Fe/H] $\le + 0.3$ 
found by Nissen \& Edvardsson (1992) is likely an artifact. Because the 
contribution of the Ni line is negligible below [Fe/H] $< -1$, our findings 
do not affect
how the O/Fe ratio changes at low metallicities and, therefore, do not resolve
the controversy mentioned in the Introduction. Nonetheless, acknowledging the
presence of the Ni\,{\sc i} blend in the solar line has the effect of 
 increasing [O/Fe] by roughly 0.13 dex in  
stars with [Fe/H] $< -1$. 

This {\it Letter} exemplifies the importance of detailed line profiles and 
accurate wavelength scales in chemical analyses of stars from spectra, 
as well as the need for hydrodynamical  model atmospheres in fine analyses of
stellar spectra.

\acknowledgments We thank Poul Nissen for inquiring about the Ni\,{\sc i} 
line's effect on oxygen abundance determinations, and both Sveneric Johansson 
and the anonymous referee for alerting us
to the reinvestigations of the Ni\,{\sc i} spectrum. This work has been 
possible with the financial help of the US National Science Foundation 
(grant AST-0086321), the Robert A. Welch Foundation of Houston (Texas), 
and the Swedish Natural Science Foundation (NFR F990/1999).

\clearpage

\end{document}